\documentclass{article} 
\usepackage[preprint]{dashboard}

\usepackage{amsmath,amsfonts,bm}

\def\eqref#1{equation~\ref{#1}}

\def\1{\bm{1}}

\DeclareMathAlphabet{\mathsfit}{\encodingdefault}{\sfdefault}{m}{sl}
\SetMathAlphabet{\mathsfit}{bold}{\encodingdefault}{\sfdefault}{bx}{n}

\usepackage{hyperref}
\usepackage{booktabs,lipsum}
\usepackage{url,comment}
\usepackage{natbib,amsthm}
\usepackage{float, caption}
\usepackage{graphicx}
\usepackage{threeparttable}
\usepackage{algorithm}
\usepackage{algorithmicx}
\usepackage{subfigure}
\usepackage{algpseudocode}

\newtheorem{remark}{Remark}
\usepackage{multirow}
\usepackage{fvextra}
\usepackage{tabularx}
\usepackage{cleveref}

\usepackage{wasysym}
\usepackage{tikz}
\usetikzlibrary{arrows.meta, positioning, shapes.misc, shapes.symbols,  decorations.pathmorphing, shapes, shapes.geometric, matrix}

\begin{document}

\title{Improving LLM Interpretability and Performance via Guided Embedding Refinement for Sequential Recommendation}

\author{
  Nanshan Jia\textsuperscript{1,2}\thanks{Nanshan Jia and Chenfei Yuan contributed equally to this work.} \quad
  Chenfei Yuan\textsuperscript{3}\footnotemark[1]\quad 
  Yuhang Wu\textsuperscript{1,2} \quad
  Zeyu Zheng\textsuperscript{1,2}\thanks{Corresponding author. Emails:\{nsjia, wuyh, zyzheng\}@berkeley.edu, yuancf21@mails.tsinghua.edu.cn} \\
  \textsuperscript{1}Department of Industrial Engineering \& Operations Research, University of California, Berkeley \\
  \textsuperscript{2} Berkeley AI Research Lab (BAIR)\\
  \textsuperscript{3}Department of Computer Science and Technology, Tsinghua University \\
}

\maketitle

\begin{abstract}
The fast development of Large Language Models (LLMs) offers growing opportunities to further improve sequential recommendation systems. Yet for some practitioners, integrating LLMs to their existing base recommendation systems raises questions about model interpretability, transparency and related safety. To partly alleviate challenges from these questions, we propose guided embedding refinement, a method that carries out a guided and interpretable usage of LLM to enhance the embeddings associated with the base recommendation system. Instead of directly using LLMs as the backbone of sequential recommendation systems, we utilize them as auxiliary tools to emulate the sales logic of recommendation and generate guided embeddings that capture domain-relevant semantic information on interpretable attributes. Benefiting from the strong generalization capabilities of the guided embedding, we construct refined embedding by using the guided embedding and reduced-dimension version of the base embedding. We then integrate the refined embedding into the recommendation module for training and inference. A range of  numerical experiments demonstrate that guided embedding is adaptable to various given existing base embedding models, and generalizes well across different recommendation tasks. The numerical results show that the refined embedding not only improves recommendation performance, achieving approximately $10\%$ to $50\%$ gains in Mean Reciprocal Rank (MRR), Recall rate, and Normalized Discounted Cumulative Gain (NDCG), but also enhances interpretability, as evidenced by case studies.
\end{abstract}

\section{Introduction}
Our work considers recommendation systems in the context of e-commerce, where good recommendations have been essential for improving customer shopping experiences \cite{resnick1997recommender,schafer1999recommender,jannach2010recommender}. In a recommendation system, items refer to the entities being recommended and users represent the main recipients of those recommendations. Interactions denote the relationships between users and items, such as viewing, purchasing, rating, or commenting. A common recommendation task involves assessing whether a specific item should be recommended to a user, based on the item's characteristics and the user's past interaction history. Among various types of recommendation systems, sequential recommendation has taken an increasingly important role in recent years. One characteristic of sequential recommendation is that a user's past interaction history can be represented in chronological order. The chronological order allows the recommendation system to flexibly assign different weights to each interaction based on its relative position in the interaction history, allowing for a more accurate analysis and a better capture of user preferences over time \cite{ezeife2023survey}.

Due to the assistance of Large Language Models (LLMs) and transformer architecture, the performance of sequential recommendation systems has been improved in two key ways. First, LLMs provide improvement for the embedding of users and items. Traditional recommendation systems rely on ID-based embeddings, where users and items are represented as unique IDs in a shared latent space \cite{goldberg1992using,sarwar2001item,koren2009matrix}. Although these embeddings are effective for calculating similarity and making recommendations, they have not adequately utilized the language-based information, especially so when a new user or item with limited interaction history enters the recommendation system \cite{sarwar2001item,kim2024large}. LLMs can utilize their power to improve the understanding of contextual information about users and items. These information are prepared as prompts and converted into embeddings, which can then be used for subsequent recommendation tasks. Second, the transformer architecture, along with its attention mechanism, has proven to be effective for sequential tasks \cite{vaswani2017attention}. By employing transformers as the backbone model for sequential recommendation, these recommendation systems achieve superior performance compared to previous methods based on Markov chains and recurrent neural networks (RNNs) \cite{kang2018self,sun2019bert4rec}.

Although the integration of LLMs has improved the performance of sequential recommendation systems, the black-box nature of LLMs raises questions about model complexity and the transparency of decision-making process. Because of this, some practitioners feel not as certain about using LLM-based sequential recommendation and they have expressed a desire for improved interpretability and explainability in the use of LLM for recommendation tasks \cite{medium_blog}. Several studies \cite{gao2023chat,lei2023recexplainer,luo2024unlocking,shi2024llm} focus on improving the explainability of LLMs in the context of recommendation systems. These approaches either leverage prompt engineering to generate explanations directly from the API \cite{gao2023chat} or fine-tune pre-trained LLMs for explanation \cite{lei2023recexplainer}. Both approaches offer contextual explanations for recommendations, thereby enhancing model explainability. They share the advantage of providing direct, word-based explanations, making them straightforward and easy to understand for users of the recommendation system.

Despite the aforementioned benefits, the existing works on improving explainability of LLM-assisted recommendations have certain challenges, which may be summarized as follows. First, these models generate explanations by producing sentences with detailed semantic content. The text generation process introduces extra computational overhead, increasing the overall inference time. When the generated explanations are relatively long, this added complexity can lead to delays in generating recommendations, which may not be appropriate for online applications or scenarios where real-time or near-instantaneous recommendations are critical. Second, some explainable recommendation models do not lead to a direct improvement in the overall recommendation performance. For example, \cite{gao2023chat} uses the OpenAI API to generate explanations, but this process is separate from the recommendation system and does not improve the quality of the recommendations. Third, many existing explainable recommendation models are difficult to generalize across different recommendation tasks. These models are often trained or fine-tuned for a specific task on a particular dataset. Although they can generate high-quality explanations within that context, they are not typically designed to generate explanations for other types of recommendation tasks. This limits their versatility, as applying them to new tasks requires retraining or fine-tuning multiple pre-trained LLMs.

In this work, we propose an embedding refinement method designed to enhance the interpretability of recommendations while preserving key advantages such as fast inference speed, improved recommendation performance, and adaptability across various recommendation tasks. Before summarizing our method, we would like to briefly discuss some of the thoughts that motivated this work. Many recommendations in the era without internet were made by experienced sales persons. For recommending clothes, an experienced sales person would evaluate a customer's purchasing attributes, such as spending budget, willingness for luxury items, preference for durability versus trendiness, brand loyalty, etc. For recommending other categories of goods, each category may have accumulated in the past some particular purchasing attributes that are useful to better assess the customer's preferences. Our work tries to explicitly use such purchasing attributes to guide the use of LLMs to generate embeddings for customers and for items. 

Our work takes a base recommendation system as given, where such base recommendation system is capable of embedding both users and items, and generating recommendations based on those embeddings; for example, SASRec \cite{kang2018self}, Bert4Rec \cite{sun2019bert4rec} and GRU4Rec \cite{hidasi2015session}. The embeddings produced by the base recommendation system are referred to as base embeddings. To enhance the interpretability and performance of the base recommendation system, we introduce the use of guided embedding, which consists of a series of scores generated by an LLM to assess domain-specific purchasing attributes. We fine-tune the LLM on a recommendation task to generate guided embeddings for both users and items. One flexibility of our method is that the recommendation task does not have to be the same with the base recommendation task, yet still can offer improvements for the based recommendation task. In our method, the guided embeddings are fine-tuned to capture how users and items align with those domain-specific, interpretable attributes. Guided embeddings trained on one task can be transferred to another, making it easy to generalize among various recommendation tasks. The refined embeddings are jointly formed by the guided embeddings and based embeddings with reduced dimension. These refined embeddings are then passed into the recommendation module for training and inference. 

Our contribution can be summarized as follows.
\begin{enumerate}
    \item We introduced the use of guided embedding, where pre-trained LLMs are fine-tuned to score items and users based on domain-specific, interpretable attributes included in prompts. The guided embedding is shown to be generalizable and easy to adapt to different recommendation tasks.
    
    \item We formed refined embedding, using guided embedding and base embedding with reduced dimensions. By incorporating refined embedding to the base recommendation module, we show improvement on both interpretability and recommendation accuracy compared to the base recommendation system. In terms of MRR, recall rate and NDCG, we achieved up to $50\%$ improvement.
    
    \item We conducted a range of experiments demonstrating that the refined embedding improves recommendation performance compared to the base embedding, even when the based embedding is of higher dimension than our refined embedding. Additionally, we illustrated the enhanced interpretability of the refined embedding through concrete case studies.
\end{enumerate}

\subsection{Literature Review}
$\bullet$ \textbf{Sequential Recommendation.} Prior to the recent advances in attention mechanisms and transformer architectures, sequential recommendation systems were broadly categorized into two main classes: Markov-chain-based system and RNN-based system. Markov-chain-based recommendation systems \cite{rendle2010factorizing,he2016fusing,he2016vista,he2017translation,tang2018personalized} assume that the next user-item interaction is determined by one or more previous interactions. RNN-based recommendation systems \cite{hidasi2015session,tan2016improved,quadrana2017personalizing,jing2017neural,donkers2017sequential,huang2018improving,hidasi2018recurrent} utilize recurrent neural networks (RNNs) to model and capture user interaction sequences. Following the advent of transformers and attention mechanisms \cite{vaswani2017attention}, the performance of sequential recommendation systems has improved greatly. These studies focus on applying transformers to capture user preferences by analyzing the sequential order of their historical interactions \cite{kang2018self,sun2019bert4rec,wu2020sse,de2021transformers4rec,li2023graph,zhang2023adaptive,chen2024explicit,liu2024probabilistic}.

\noindent$\bullet$ \textbf{LLM for Recommendation.} 
Before the advent of LLMs, recommendation systems primarily relied on collaborative filtering \cite{sarwar2001item,ahn2008new,sun2011novel,bobadilla2012collaborative} or matrix factorization techniques \cite{koren2009matrix}. These ID-based methods faced challenges with the cold-start problem, where users and items with limited interaction histories were difficult to integrate into the system. To address this issue, LLM-based recommendation systems leverage the contextual capabilities of LLMs to generate embeddings for items based on their contextual descriptions and for users based on their past interactions \cite{sanner2023large,liu2024large,chen2024hllm,kim2024large,zhai2024actions,wu2024coral,lin2024bridging,wang2024warming}. In addition to the embedding approach, there are a series of work using prompt engineering to improve the recommendations \cite{penha2020does,zhang2023prompt,lyu2023llm,li2023prompt,liu2023recprompt}. Multi-modal information are also incorporated into recommendation systems to enhance the recommendation performance \cite{liu2024rec,tian2024mmrec,wang2024mllm4rec,ye2024harnessing,zhang2024notellm,xv2024improving}.

\noindent$\bullet$ \textbf{LLM for Recommendation Explanation.}
The explainability of LLMs is defined by \cite{zhao2024explainability} as \textit{the ability to explain or present the behavior of models in human-understandable terms}. Enhanced explainability not only fosters greater trust in system decisions but also enables model improvement through explanations. In the context of recommendation systems, there are several main approaches. The API approach generates explanations for recommendations by directly utilizing publicly available APIs. These methods employ prompt engineering techniques or build conversational systems to elicit contextual explanations \cite{gao2023chat,zhou2023gpt,liu2023chatgpt,wang2023improving,li2023gpt4rec,feng2023large,lubos2024llm}. The surrogate model approach typically involves using a pre-trained LLM as a backbone surrogate model. This surrogate model is further trained or fine-tuned to replicate the performance of a recommendation system while simultaneously generating explanations \cite{lei2023recexplainer,li2023ucepic,bismay2024reasoningrec}. The prompt approach focuses on refining prompts to uncover personalized information and generate explanations \cite{li2023personalized,li2023prompt}. Reinforcement learning has also been explored to enhance model explainability and performance \cite{xue2023prefrec,zhang2024natural}.

\section{Problem Formulation}
Let $\mathcal{I}$ denote the set of items in a recommendation system, where each item $i\in\mathcal{I}$ has a unique text description $v_i$, such as its name, title, or properties. Let $\mathcal{U}$ represent the set of users, with each user having multiple interactions with a subset of items. Specifically, for a user $u \in \mathcal{U}$, its past interactions are represented as a sequence $S_u = \{(i_1, f_1), (i_2, f_2), \dots, (i_n, f_n)\}$ ordered chronologically. Here, $i_1, \dots, i_n$ are the items that the user $u$ has interacted with. Each pair of interactions $(u, i_k)$ is associated with a feedback $f_k$, which may include a rating, a comment, or an attitude towards the item $i_k$. The goal of sequential recommendation is to predict the user's next interaction from a set of possible candidate items. Specifically, the task is to predict the next item $i_{n+1}$ as follows:
\begin{equation}
i_{n+1} = \underset{i \in \mathcal{I}_u}{\arg\max} \;\mathbb{P}(i \mid S_u),
\end{equation}
where $\mathcal{I}_u \subset \mathcal{I}$ denotes the set of candidate items for user $u$. 

A base recommendation system consists of two key components: the embedding module and the recommendation module. The embedding module transforms the text description $v_i$ for each item $i \in \mathcal{I}$ into an embedding $e_i \in \mathbb{R}^d$. For each user $u \in \mathcal{U}$ with past interactions $S_u$, the embedding module learns to map the user $u$ to an embedding $e_u\in\mathbb{R}^d$, which resides in the same latent space as item embeddings and aims to capture the user preferences through the interaction history $S_u$. The recommendation module takes the user embedding $e_u$ and the embeddings of the items in the candidate set $\mathcal{I}_u$ as input and outputs the item with the highest probability of being recommended as the next interaction item. For convenience, we denote the embedding module as $\mathcal{E} : \mathcal{I} \cup \mathcal{U} \to \mathbb{R}^d$ and the recommendation module as $\mathcal{R} : \{(e_u, \{e_i:i\in \mathcal{I}_u\}) : u \in \mathcal{U}\} \to \mathcal{I}$. During training, the objective is to maximize the likelihood of correctly predicting the next item. To achieve this, the recommendation system typically minimizes a loss function, such as the cross-entropy loss, between the predicted item and the true item. Without loss of generality, we denote the loss function of the base recommendation system as $\mathcal{L}_{\mathrm{base}}$, i.e.,
\begin{equation}
    \mathcal{L}_{\mathrm{base}}=-\sum_{u\in\mathcal{U}}l(i_{n+1}\mid e_u,\mathcal{I}_u),\label{loss;base}
\end{equation}
where $e_u$ is the user embedding, $i_{n+1}$ is the ground-truth next item and $l$ is a function positively correlated to the probability mass function. For example, $l(\cdot\mid e_u,\mathcal{I}_u)=\log \mathbb{P}(\cdot\mid e_u,\mathcal{I}_u)$ if the system uses cross-entropy loss.

\section{Our Method}
In this section, we introduce our guided embedding refinement method. Specifically, we propose the guided embedding for both items and users in Section \ref{guided embedding}. The combination of guided embedding and base embedding to form the refined embedding is discussed in Section \ref{refined embedding}. Throughout this section, we assume that a pre-trained LLM, denoted as $f$, is always capable of providing an output when it receives prompt-based input.

\subsection{Fine-tuning LLMs for Guided Embedding}
\label{guided embedding}
Suppose $a_1, a_2, \dots, a_m$ are $m$ interpretable, domain-specific aspects of the items and are described in a given contextual format. For example, $a_1$ could be "tension level" if the items are movies. For each item $i\in\mathcal{I}$, we query the pre-trained large language model (LLM) to score the item according to these $m$ aspects via zero-shot prompt. The prompt contains both the scoring task and the regulation of the output format. This process yields the guided embedding for the item, defined as:
\begin{equation}
e_i^g := h\circ f(v_i\mid\mathrm{prompt},a_1,\cdots,a_m),
\end{equation}
where $ v_i $ is the text description, $h$ is a function designed for extracting scores from the output of LLM $f$ and $e_i^g\in\mathbb{R}^{m}$ is the guided embedding of item $i$. The prompt format and specific aspect choices are listed in Appendix \ref{prompt format}.

Similarly, the guided embedding for a given user $u\in\mathcal{U}$ can be derived by 
\begin{equation}
    e_u^g:=h\circ f(S_u\mid\mathrm{prompt},a_1,\cdots,a_m).
\end{equation}
In practice, we freeze the guided item embeddings during training, i.e., the guided embeddings for items remain unchanged throughout the training process, while the guided user embeddings are trainable. 

Next, we conduct a detailed recommendation task and fine-tune the pre-trained LLM for guided user embeddings using LoRA \cite{hu2021lora}. The recommendation task can be one of the following:
\begin{enumerate}
    \item \textbf{Base task.} The same recommendation task as the base recommendation system.
    \item \textbf{Classification task.} Predicting whether or not a user likes an item.
    \item \textbf{Other recommendation tasks that the system operator prefers.}
\end{enumerate}
The loss function for fine-tuning LLM consists of two main components: the recommendation loss and the format loss. The recommendation loss corresponds to the loss function specific to the recommendation task. For example, if we consider the same recommendation task as the base recommendation system, the recommendation loss can be defined as
\begin{equation}
    \mathcal{L}_{\mathrm{rec}}=-\sum_{u\in\mathcal{U}}l(i_{n+1}\mid e_u^g,\mathcal{I}_u).
\end{equation}
Alternatively, for the classification task, the recommendation loss would reflect the number of misclassified user-item interaction pairs.

In addition, as mentioned earlier, the LLM $f$ is supposed to provide output in a certain format specified in the input prompt. Hence, we add a format loss, which is defined as cross-entropy loss, to regulate the output format:
\begin{equation}
    \mathcal{L}_{\mathrm{format}}=-\sum_{u\in\mathcal{U}}\log\mathbb{P}(\mathrm{mask}\circ f(\cdot)\mid\mathrm{prompt,a_1,\cdots,a_m}),
\end{equation}
where $f(\cdot)$ is the LLM output and $\mathrm{mask}$ is the mask operation that masks the scores. The mask operation guarantees that the format loss is not influenced by the scores. In total, the loss function for fine-tuning $f$ is given by
\begin{equation}
\mathcal{L}_{ge}=\mathcal{L}_{\mathrm{rec}}+\lambda \mathcal{L}_{\mathrm{format}},
\end{equation}
where $\lambda>0$ is a hyperparameter controlling the weight of format loss in the overall loss. 

\begin{remark}
As discussed in this section, the recommendation task for fine-tuning the pre-trained LLM does not have to match the primary sequential recommendation task of the model. Instead, it can be any recommendation task, even as simple as a classification task, and is selected based on the system operator's preferences. The key point behind the guided embeddings is to capture how items and users align with those domain-specific, interpretable attributes. The choice of the recommendation task itself is not central to this part. Guided embeddings trained on one task can be transferred and applied to other tasks, making our method highly generalizable. This overcomes the task-specific limitations of previous interpretable models, providing a more flexible solution for various recommendation scenarios. We will further discuss this benefit in Section \ref{Effect of Recommendation Task on Guided Embedding Generation} later.
\end{remark}

\begin{figure}[!htbp]
    \centering
    {\includegraphics[width=0.6\textwidth]{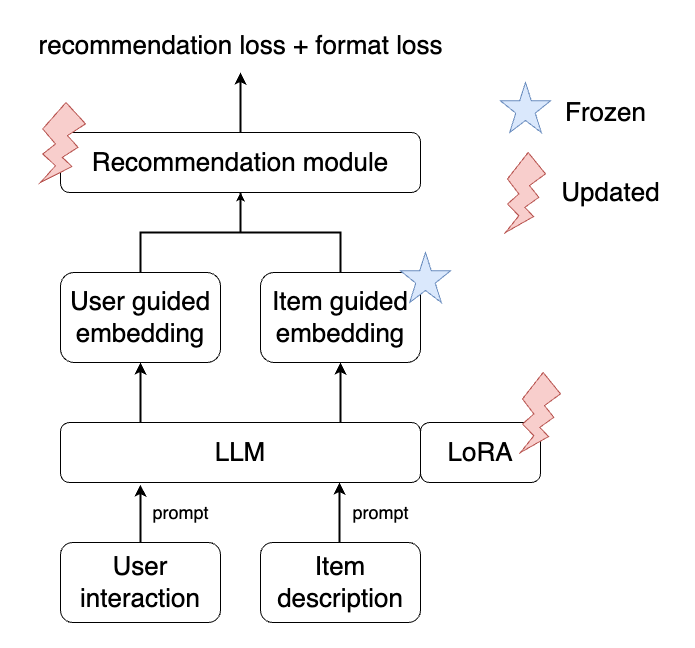}}
    \caption{Process for fine-tuning the LLM for guided embeddings. The recommendation module can be any among base recommendation module, binary classification module, or other related recommendation tasks that the system operator prefers.}
    \label{model1}
\end{figure}

Before concluding this section, we summarize the process for fine-tuning the guided embedding model in Figure \ref{model1}.

\subsection{Integrating Guided Embedding to the Base Recommendation System}
\label{refined embedding}
\subsubsection{Normalization}
Before passing the guided embeddings to the base recommendation system, we apply a normalization step. This is crucial because the raw scores provided by the LLM fall within the range of $1$ to $10$. Such a range can lead to performance issues, as the mean and variance of these scores are relatively large, causing the guided embedding to dominate the recommendation module's learning process. To address this problem, we apply a linear transformation to normalize the guided embeddings:
\begin{subequations}
\begin{align}
    e_i^g &\leftarrow W e_i^g + B, \\
    e_u^g &\leftarrow W e_u^g + B,
\end{align}
\end{subequations}
where $W\in\mathbb{R}^{m\times m}$ is a weight matrix and $B\in\mathbb{R}^m$ is a bias term. After normalization, the guided embedding is sent to the base recommendation system for further training and inference.

\subsubsection{Refined Embedding}
To avoid ambiguity, we use superscript $b$ to denote the base embeddings. We define the refined embeddings by
\begin{subequations}
\begin{align}
e_i^r &=e_i^b \oplus (\mu * e_i^g), \\
e_u^r &=e_u^b \oplus (\mu * e_u^g), 
\end{align}
\end{subequations}
where $\oplus$ represents concatenation and $\mu>0$ is a hyperparameter controlling the weight of the guided embedding in the refined embedding. 

\subsubsection{Sequential Recommendation}
\begin{figure}[!htbp]
    \centering
    {\includegraphics[width=0.7\textwidth]{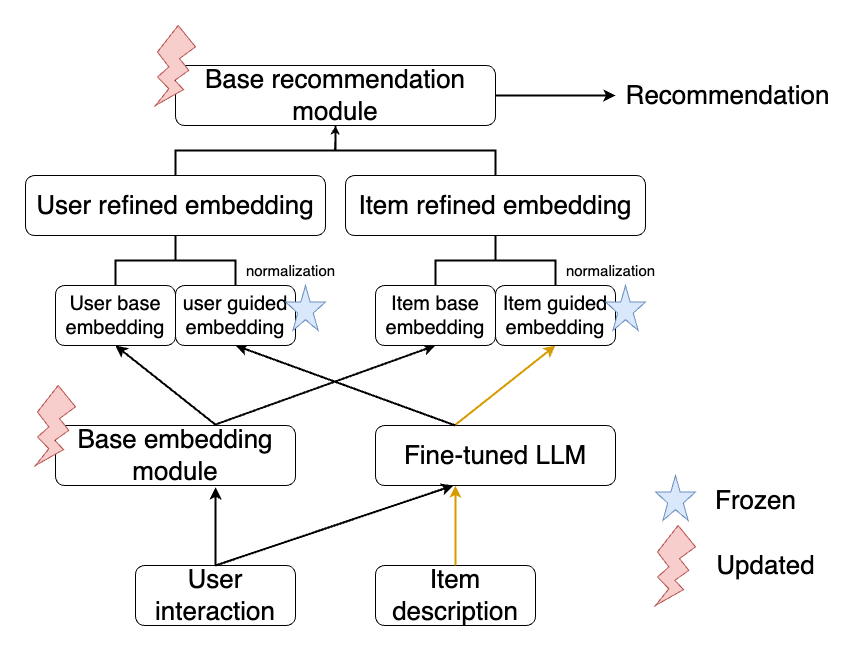}}
    \caption{Process for integrating guided embedding to the base recommendation system.}
    \label{model2}
\end{figure}

To train the sequential recommendation module and the base embeddings, we adopt the following modified loss:
\begin{equation}
    \mathcal{L}_{\mathrm{ref}}=-\sum_{u\in\mathcal{U}}l(i_{n+1}\mid e_u^r,\mathcal{I}_u).
\end{equation}
Here, the guided embedding component is frozen within the refined embedding. Compared to (\ref{loss;base}), the key distinction is that the guided embedding is included in the loss function. As a result, the base embeddings in the current model differ from those in the original base recommendation system. Indeed, the guided embeddings are primarily responsible for reflecting the performance of items and users across the pre-selected attributes, while the base embeddings handle the specific downstream recommendation task. We summarize the process of integrating the guided embedding into the base recommendation system in Figure \ref{model2}.

\vspace{-0.3cm}

\vspace{-0.3cm}

\section{Numerical Experiments}
\label{Numerical Experiments}
\subsection{Datasets}
We conducted our experiments using the publicly available Amazon Reviews 2023 dataset, collected by \cite{hou2024bridging}.
This dataset comprises user reviews, item metadata, and interaction records from Amazon, covering from May 1996 to September 2023. To evaluate the effectiveness of our guided embedding refinement method, we select three different product categories: 1) \textbf{Movies} and TV; 2) \textbf{Clothing}, Shoes, and Jewelry; and 3) Video \textbf{Games}. By selecting these three distinct domains, we aim to assess the reliability of our approach across different categories of the sequential recommendation tasks. For data cleaning, we have followed the standard data pre-processing pipeline in the literature, and the details can be found in Appendix \ref{data;prep}.

The key statistics of the processed datasets for different categories are presented in \autoref{tab:ds_desc}.
\begin{table}[!htbp]
    \centering
    \begin{tabular}{lrrr}
        \toprule
        Category & \#users & \#items & \#reviews \\
        \midrule
        Clothing & 4472 & 16101 & 228730 \\
        Movies & 1200 & 4357 & 99185 \\
        Games & 370 & 555 & 12255 \\
        \bottomrule
    \end{tabular}
    \caption{Statistics of datasets}
    \label{tab:ds_desc}
\end{table}

\subsection{Experimental Setup}

\subsubsection{Guided Embedding Generation}
\label{guided emb gen}
For guided embedding generation, we use the pre-trained Llama 3.2 1B model as the backbone model and fine-tune it for a recommendation task based on LoRA \cite{hu2021lora}. 
The fine-tuning process involves minimizing the recommendation loss and the format loss, as described in Section~\ref{guided embedding}.
During fine-tuning, the hyperparameter $\lambda$ is set to $0.1$ if the model successfully extracts guided embeddings; and it is set to $1$ otherwise to prioritize output formatting.
If structured output is missing, $\mathcal{L}_{\mathrm{rec}}$ is set to $0$ to prevent training instability.
The recommendation loss varies according to the recommendation task:
(1) \textbf{Base task}: mirrors the training objective of the base recommendation model. 
For instance, in BERT4Rec~\cite{sun2019bert4rec}, the model is trained to predict the next item in a user’s interaction sequence.
We use the same objective for fine-tuning the LLM, and evaluate its performance using Mean Reciprocal Rank (MRR); and
(2) \textbf{Classification task}: a binary classification task in which the model assesses whether a user \textit{likes} an item based on historical interactions.
Items rated above a predefined threshold are considered \textit{liked}, while all others are not.
We evaluate the model using the Area Under the Receiver Operating Characteristic Curve (AUC-ROC).

To generate guided embeddings, we design domain-specific prompts that instruct the LLM to score each item along multiple interpretable attributes.
For example, in the Movies and TV category, the aspects include "story complexity", "dialogue complexity", "intellectual challenge", etc.
The prompt format for generating guided embeddings is tailored to each category, with the specific attributes listed in \autoref{prompt format}.
Each attribute is scored on a 1-10 scale, and the LLM is prompted to generate responses in a structured format.
We then apply a rule-based extraction method to parse these scores, forming the guided embedding $e_i^g$ for each item $i$. 
Similarly, for user embeddings, we prompt the LLM to analyze a user's historical interactions and generate scores reflecting user preferences across the same predefined aspects. 
Unlike item guided embeddings, which is frozen, user guided embeddings remain trainable, allowing them to dynamically align with the item guided embeddings on those attributes.

\subsubsection{Base Recommendation System}
To verify the effectiveness and generalizability of our guided embedding refinement method, we select two RNN-based recommendation systems, GRU4Rec \cite{hidasi2015session} and GRU4Rec$^+$ \cite{hidasi2015session}, as well as two transformer-based recommendation systems, SASRec \cite{kang2018self} and BERT4Rec \cite{sun2019bert4rec}, as our base recommendation systems. More information of these four models can found in Appendix \ref{app;base}. 
For training the base recommendation models, we adopt the original hyperparameter settings specified in the respective papers.

\subsubsection{Evaluation Metrics}
In this work, we primarily use three evaluation metrics to compare the sequential recommendation performance between refined embedding and base embedding.
\begin{itemize}
    \item \textbf{Maximum Marginal Relevance (MMR).} MMR is a ranking strategy introduced to balance the trade-off between relevance and diversity in recommendation systems \cite{carbonell1998use}. A larger MMR indicates stronger relevance and reduced redunduncy in recommendation.
    \item \textbf{Recall@k.} Recall@k is the proportion of the top-k recommendations that include the ground-truth item. It measures how well the system retrieves the ground-truth items, regardless of their rank or diversity. This metric emphasizes coverage over diversity.
    \item \textbf{Normalized Discounted Cumulative Gain (NDCG@k).} In this metric, we define the relevance of an item as 1 if it is the ground-truth and 0 otherwise. NDCG@k is calculated based on the top-k recommendations and focuses on ranking relevant items higher, penalizing irrelevant ones based on their positions in the ranked list \cite{wang2013theoretical}. A higher NDCG score (closer to 1) indicates better ranking quality in terms of relevance.
\end{itemize}

\subsection{Performance Comparison}
\begin{table*}[!htbp]
\tiny
\centering
    \begin{tabular}{lccrccrccrccr}
    \toprule
     \multirow{2}{*}{Metric} 
        & \multicolumn{3}{c}{GRU4Rec} 
        & \multicolumn{3}{c}{GRU4Rec$^+$} 
        & \multicolumn{3}{c}{SASRec} 
        & \multicolumn{3}{c}{BERT4Rec}  \\
        \cmidrule(lr){2-4}\cmidrule(lr){5-7}\cmidrule(lr){8-10}\cmidrule(lr){11-13}
          & Base. & Ref. & Imp. & Base. & Ref. & Imp. & Base. & Ref. & Imp. & Base. & Ref. & Imp. \\
        \midrule
          MRR & 0.0148 & 0.0159 & 7.80\% & 0.0105 & 0.0116 & 11.10\% & 0.2054 & \textbf{0.2227} & 8.44\% & 0.1995 & 0.2166 & 8.54\% \\
         Recall@1 & 0.0076 & 0.0085 & 11.34\% & 0.0042 & 0.0059 & 40.10\% & 0.1025 & \textbf{0.1144} & 11.65\% & 0.0925 & 0.1040 & 12.44\% \\
         Recall@5 & 0.0179 & 0.0190 & 6.11\% & 0.0144 & 0.0150 & 4.17\% & 0.2883 & \textbf{0.3154} & 9.40\% & 0.2874 & 0.3124 & 8.69\% \\
        Recall@10 & 0.0251 & 0.0265 & 5.39\% & 0.0205 & 0.0211 & 3.17\% & 0.4155 & 0.4549 & 9.49\% & 0.4301 & \textbf{0.4611} & 7.21\% \\
         Recall@20 & 0.0348 & 0.0367 & 5.60\% & 0.0277 & 0.0280 & 1.01\% & 0.5818 & 0.6166 & 5.98\% & 0.5994 & \textbf{0.6298} & 5.06\% \\
         NDCG@5 & 0.0129 & 0.0138 & 7.53\% & 0.0094 & 0.0104 & 10.57\% & 0.1968 & \textbf{0.2160} & 9.75\% & 0.1906 & 0.2095 & 9.90\% \\
         NDCG@10 & 0.0152 & 0.0162 & 6.94\% & 0.0114 & 0.0124 & 8.77\% & 0.2378 & \textbf{0.2610} & 9.76\% & 0.2366 & 0.2574 & 8.79\% \\
         NDCG@20 & 0.0176 & 0.0188 & 6.84\% & 0.0132 & 0.0141 & 6.85\% & 0.2797 & \textbf{0.3018} & 7.90\% & 0.2793 & 0.3000 & 7.43\% \\
        \bottomrule
    \end{tabular}
    \begin{flushleft}
        \textit{Note that the total dimensionality of the base embedding and refined embedding remains the \textbf{same}.}
    \end{flushleft}
    \caption{Performance comparison between base and refined embeddings across different recommendation models on the Clothing dataset.}
    \label{tab:performance;1}
\end{table*}

\begin{table*}[!htbp]
\tiny
\centering
    \begin{tabular}{lccrccrccrccr}
    \toprule
     \multirow{2}{*}{Metric} 
        & \multicolumn{3}{c}{GRU4Rec} 
        & \multicolumn{3}{c}{GRU4Rec$^+$} 
        & \multicolumn{3}{c}{SASRec} 
        & \multicolumn{3}{c}{BERT4Rec}  \\
        \cmidrule(lr){2-4}\cmidrule(lr){5-7}\cmidrule(lr){8-10}\cmidrule(lr){11-13}
          & Base. & Ref. & Imp. & Base. & Ref. & Imp. & Base. & Ref. & Imp. & Base. & Ref. & Imp. \\
        \midrule
          MRR & 0.0244 & 0.0263 & 7.47\% & 0.0222 & 0.0237 & 6.63\% & 0.1939 & \textbf{0.2140} & 10.38\% & 0.1648 & 0.1880 & 14.05\% \\
         Recall@1 & 0.0120 & 0.0130 & 8.17\% & 0.0112 & 0.0121 & 8.81\% & 0.1012 & \textbf{0.1187} & 17.30\% & 0.0730 & 0.0952 & 30.41\% \\
         Recall@5 & 0.0299 & 0.0325 & 8.58\% & 0.0275 & 0.0295 & 7.22\% & 0.2648 & \textbf{0.2902} & 9.57\% & 0.2280 & 0.2551 & 11.85\% \\
         Recall@10 & 0.0429 & 0.0462 & 7.63\% & 0.0401 & 0.0421 & 4.98\% & 0.3677 & \textbf{0.3922} & 6.66\% & 0.3398 & 0.3765 & 10.79\% \\
         Recall@20 & 0.0614 & 0.0661 & 7.74\% & 0.0558 & 0.0582 & 4.41\% & 0.4992 & \textbf{0.5302} & 6.21\% & 0.4931 & 0.5208 & 5.62\% \\
         NDCG@5 & 0.0212 & 0.0229 & 8.35\% & 0.0194 & 0.0209 & 7.58\% & 0.1859 & \textbf{0.2076} & 11.65\% & 0.1525 & 0.1766 & 15.85\% \\
         NDCG@10 & 0.0253 & 0.0273 & 7.89\% & 0.0235 & 0.0250 & 6.31\% & 0.2190 & \textbf{0.2404} & 9.80\% & 0.1884 & 0.2156 & 14.42\% \\
         NDCG@20 & 0.0300 & 0.0323 & 7.87\% & 0.0274 & 0.0290 & 5.77\% & 0.2520 & \textbf{0.2750} & 9.10\% & 0.2270 & 0.2520 & 11.03\% \\
        \bottomrule
    \end{tabular}
    \begin{flushleft}
        \textit{Note that the total dimensionality of the base embedding and refined embedding remains the \textbf{same}.}
    \end{flushleft}
    \caption{Performance comparison between base and refined embeddings across different recommendation models on the Movies dataset.}
    \label{tab:performance;2}
\end{table*}

\begin{table*}[!htbp]
\tiny
\centering
    \begin{tabular}{lccrccrccrccr}
    \toprule
     \multirow{2}{*}{Metric} 
        & \multicolumn{3}{c}{GRU4Rec} 
        & \multicolumn{3}{c}{GRU4Rec$^+$} 
        & \multicolumn{3}{c}{SASRec} 
        & \multicolumn{3}{c}{BERT4Rec}  \\
        \cmidrule(lr){2-4}\cmidrule(lr){5-7}\cmidrule(lr){8-10}\cmidrule(lr){11-13}
          & Base. & Ref. & Imp. & Base. & Ref. & Imp. & Base. & Ref. & Imp. & Base. & Ref. & Imp. \\
        \midrule
         MRR & 0.0485 & 0.0576 & 18.70\% & 0.0453 & 0.0527 & 16.40\% & 0.2014 & \textbf{0.2304} & 14.39\% & 0.1544 & 0.1849 & 19.74\% \\
         Recall@1 & 0.0164 & 0.0215 & 31.30\% & 0.0138 & 0.0206 & 48.97\% & 0.0946 & \textbf{0.1184} & 25.14\% & 0.0572 & 0.0792 & 38.29\% \\
         Recall@5 & 0.0603 & 0.0760 & 25.86\% & 0.0592 & 0.0673 & 13.60\% & 0.2897 & \textbf{0.3286} & 13.43\% & 0.2201 & 0.2827 & 28.43\% \\
         Recall@10 & 0.0980 & 0.1149 & 17.24\% & 0.0944 & 0.1034 & 9.51\% & 0.4265 & \textbf{0.4643} & 8.87\% & 0.3539 & 0.4261 & 20.43\% \\
         Recall@20 & 0.1550 & 0.1750 & 12.87\% & 0.1437 & 0.1573 & 9.47\% & 0.6005 & \textbf{0.6200} & 3.24\% & 0.5418 & 0.5867 & 8.29\% \\
         NDCG@5 & 0.0384 & 0.0490 & 27.47\% & 0.0368 & 0.0441 & 19.92\% & 0.1934 & \textbf{0.2262} & 17.00\% & 0.1393 & 0.1792 & 28.65\% \\
         NDCG@10 & 0.0506 & 0.0614 & 21.44\% & 0.0481 & 0.0557 & 15.81\% & 0.2376 & \textbf{0.2697} & 13.53\% & 0.1823 & 0.2253 & 23.58\% \\
         NDCG@20 & 0.0649 & 0.0765 & 17.94\% & 0.0606 & 0.0693 & 14.36\% & 0.2815 & \textbf{0.3094} & 9.89\% & 0.2296 & 0.2657 & 15.70\% \\
        \bottomrule
    \end{tabular}
    \begin{flushleft}
        \textit{Note that the total dimensionality of the base embedding and refined embedding remains the \textbf{same}.}
    \end{flushleft}
    \caption{Performance comparison between base and refined embeddings across different recommendation models on the Games dataset.}
    \label{tab:performance;3}
\end{table*}

For each experimental setting, we conduct five independent runs and each run shows similar relative performances. We report the experiment results in \autoref{tab:performance;1}-\autoref{tab:performance;3}. Among all the tables, we adopt the following abbreviation for simplicity: "Base." and "Ref." denote the performance of the base and refined embeddings, respectively. "Imp." represents the improvement of the refined embedding over the base embedding.
Additionally, for all base recommendation models, the NDCG@1 metric is equivalent to Recall@1 under their original configurations. Therefore, we report only Recall@1 for clarity and conciseness.
The results consistently show that, among all datasets and base recommendation models, refined embeddings consistently improve recommendation performance. 
The enhancement is particularly noticeable in transformer-based models such as SASRec and BERT4Rec, where the interpretability and semantic richness of guided embeddings contribute to better user-item matching. 
The improvements in Recall@1, and NDCG@5 indicate that our method significantly benefits top-ranked recommendations, which are crucial in practical recommendation systems.

For RNN-based models, the improvements achieved with refined embedding are relatively moderate compared to those seen in transformer-based methods. This can mainly be attributed to the short-term user behavior modeling employed in GRU4Rec and GRU4Rec$^+$. The focus on short-term interactions may limit the recommendation system’s ability to leverage guided embeddings derived from the long-term interaction history to analyze user preferences.
Nevertheless, the incorporation of guided embeddings still yields significant performance gains.

\subsection{Ablation Study}
To further analyze the effectiveness of guided embedding refinement, we conduct a series of ablation studies to examine the impact of some hyperparameters on recommendation performance. Specifically, we investigate four critical aspects: 
\begin{enumerate}
    \item The dimensionality of the guided embedding.
    \item Assessing whether the superior performance of the refined embedding (60 dimensional) can be reached by a higher-dimensional (e.g., 240 dimensional) base embedding. This verifies the critical value of the refined embedding and its guided component.    
    \item The impact of different recommendation tasks used for fine-tuning the LLM when generating guided embeddings.
    \item The parameter $\mu$ controlling the weight of the guided embedding in the refined embedding.
\end{enumerate}

Due to the limited space here, we present both experiment results and analysis of the first three ablation studies in this section. The last ablation study is included in Appendix \ref{app;abl}.

\subsubsection{Effect of Guided Embedding Dimensionality}
\label{sec:int_dim}

\begin{figure}[htbp]
    \centering
    {\includegraphics[width=0.9\textwidth]{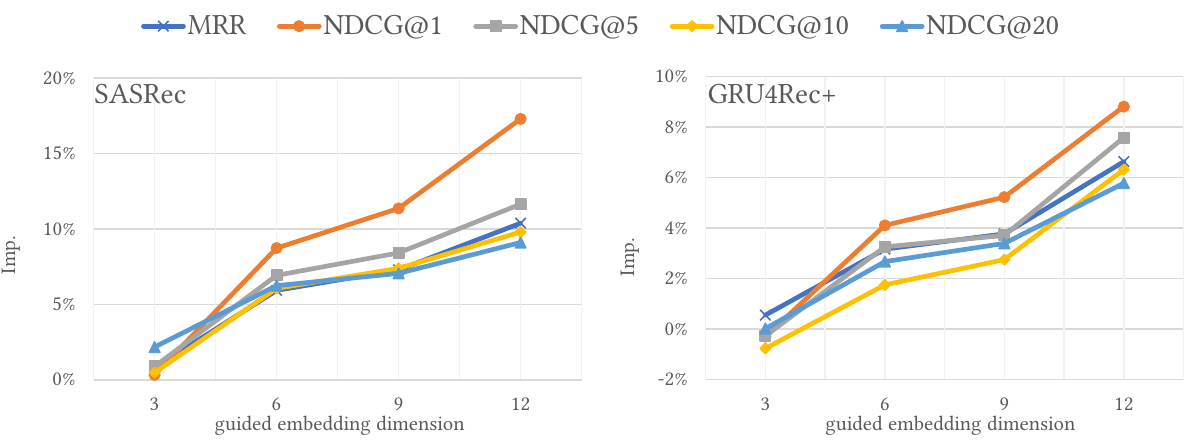}}
    \caption{Impact of guided embedding dimensionality on performance improvement for SASRec and GRU4Rec$^+$ on the Movies dataset. The x-axis represents different values of guided embedding dimensionality, and the y-axis indicates the percentage improvement in performance metrics.}
    \label{fig:int_dim}
\end{figure}

To analyze the impact of guided embedding dimensionality on recommendation performance, we experiment with different dimensions: 3, 6, 9, and 12. We evaluated the performance of SASRec, a representative transformer-based model, and GRU4Rec$^+$, a representative RNN-based model, on the Movies dataset and measured the improvement over their respective base recommendation models among various ranking metrics (MRR, NDCG@1, NDCG@5, NDCG@10, NDCG@20).

\Cref{fig:int_dim} shows that our method consistently provides significant performance improvements across \textbf{all} tested dimensions, except when the guided embedding dimension is set to 3, where the gains are minimal. In addition, the recommendation performance improves as the dimensionality of the guided embedding increases from $3$ to $12$. These results indicate a tradeoff between recommendation performance and computational efforts. As the dimensionality of the guided embedding increases, the recommendation performance is enhanced at the cost of higher computational costs.

\begin{figure*}[!htbp]
    \centering
    {\includegraphics[width=0.9\textwidth]{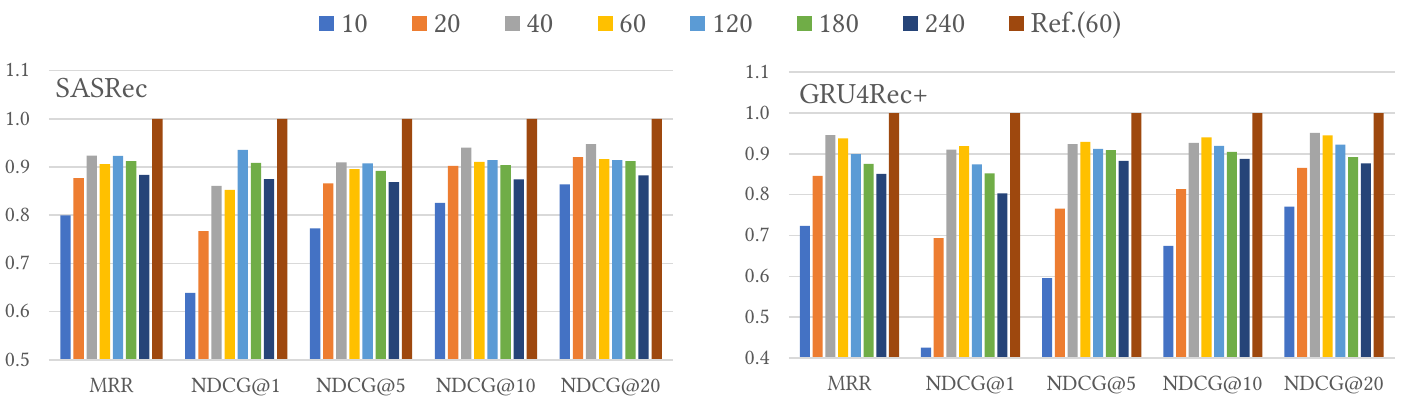}}
    \caption{Impact of base embedding dimensionality on performance for SASRec and GRU4Rec$^+$ on the Movies dataset. "Ref.(60)" represents the refined embedding setup with base embedding of size 48 combined with a guided embedding of size 12. All results are normalized with Ref.(60) set to 1.}
    \label{fig:base_dim}
\end{figure*}

\subsubsection{Assessing the Critical Gain by Guided Embedding}
To evaluate the performance gain provided by guided embedding, we conducted experiments with base embeddings of varying dimensionality. These experiments were performed on a movie dataset using SASRec and GRU4Rec$^+$. Specifically, we tested base embeddings with dimensions of 10, 20, 40, 60, 120, 180, and 240. The refined embedding (Ref.(60)) is formed by using a base embedding of 48 dimensions and a guided embedding of 12 dimensions, resulting in a total embedding size of 60. For each configuration, we select the best-performing results across different epochs and normalize the results so that Ref.(60) has a unit value. The results shown in \autoref{fig:base_dim} indicate that the refined embedding outperforms all base embeddings, even when the base embedding's dimensionality is significantly higher (up to four times larger) than that of the refined embedding.

Another interesting observation is that, as the base embedding dimension increases from 10 to 60, both MRR and NDCG scores show significant improvements. However, the performance gains begin to plateau once the dimension exceeds 60, and in some cases, a slight degradation is observed. This pattern aligns with the findings of \cite{kang2018self,sun2019bert4rec}. In other words, simply scaling up the dimension of the base embeddings cannot reach the superior performance of the refined embeddings. The performance gains of the refined embeddings comes from the domain-relevant, interpretable guided embeddings, rather than the 12 dimensions of the guided embeddings themselves.

\subsubsection{Effect of Recommendation Task on Guided Embedding Generation}
\label{Effect of Recommendation Task on Guided Embedding Generation}
\begin{table*}[!htbp]
    \centering
    \tiny
    \begin{tabular}{lcccrrcccrr}
        \toprule
         \multirow{2}{*}{Metric} 
        & \multicolumn{5}{c}{GRU4Rec$^+$} 
        & \multicolumn{5}{c}{SASRec}  \\
        \cmidrule(lr){2-6}\cmidrule(lr){7-11}
      & Base. & Ref.(base) & Ref.(cls) & Imp.(base) & Imp.(cls) & Base. & Ref.(base) & Ref.(cls) & Imp.(base) & Imp.(cls) \\
        \midrule
         MRR & 0.0222 & 0.0237 & 0.0235 & 6.63\% & 5.59\% & 0.1939 & 0.2140 & 0.2103 & 10.38\% & 8.49\% \\
         Recall@1 & 0.0112 & 0.0121 & 0.0120 & 8.81\% & 7.09\% & 0.1012 & 0.1187 & 0.1167 & 17.30\% & 15.32\% \\
         Recall@5 & 0.0275 & 0.0295 & 0.0293 & 7.22\% & 6.61\% & 0.2648 & 0.2902 & 0.2847 & 9.57\% & 7.49\% \\
         Recall@10 & 0.0401 & 0.0421 & 0.0416 & 4.98\% & 3.68\% & 0.3677 & 0.3922 & 0.3923 & 6.66\% & 6.71\% \\
         Recall@20 & 0.0558 & 0.0582 & 0.0578 & 4.41\% & 3.63\% & 0.4992 & 0.5302 & 0.5215 & 6.21\% & 4.47\% \\
         NDCG@5 & 0.0194 & 0.0209 & 0.0207 & 7.58\% & 6.59\% & 0.1859 & 0.2076 & 0.2031 & 11.65\% & 9.24\% \\
         NDCG@10 & 0.0235 & 0.0250 & 0.0247 & 6.31\% & 5.00\% & 0.2190 & 0.2404 & 0.2378 & 9.80\% & 8.62\% \\
         NDCG@20 & 0.0274 & 0.0290 & 0.0287 & 5.77\% & 4.80\% & 0.2520 & 0.2750 & 0.2704 & 9.10\% & 7.29\% \\
        \bottomrule
    \end{tabular}
     \begin{flushleft}
        \textit{Note that the total dimensionality of the base embedding and refined embedding remains the \textbf{same}.}
    \end{flushleft}
    \caption{Performance comparison across different recommendation models on the Movies dataset. "Base." represents the performance of the base recommendation model without guided embeddings. "Ref.(base)" and "Ref.(cls)" refer to refined embeddings generated from fine-tuning on the base task and the classification task, respectively. "Imp." indicates the relative improvement in performance over the base model.}
    \label{tab:rec_task}
\end{table*}

The choice of recommendation task for fine-tuning the LLM may have an impact on the effectiveness of guided embedding refinement. As discussed in Section~\ref{guided emb gen}, we explore two choices: the base task, which aligns with the recommendation task in the base recommendation system, and the binary classification task, which focuses on predicting whether a user will like an item. We conducted experiments using GRU4Rec$^+$ and SASRec as base recommendation systems on the movie dataset.

The results, presented in \autoref{tab:rec_task}, compare the base recommendation model (Base.) with refined embeddings trained using the base task (Ref.(base)) and the classification task (Ref.(cls)). Our observations show that both approaches significantly improve performance over the base recommendation system, demonstrating that guided embeddings can provide substantial performance gains regardless of the recommendation task used for fine-tuning. In other words, the specific recommendation task for fine-tuning is not critical for generating effective guided embeddings. A guided embedding fine-tuned under one task (e.g., the binary classification task) can still improve the performance of another recommendation task (e.g., the base sequential recommendation task). These findings further highlight that guided embeddings are generalizable and adaptable to different recommendation tasks.

Additionally, we note that the refined embedding trained on the base task achieves slightly better performance than the one trained on the binary classification task. This subtle improvement may come from the alignment of the base task with the final recommendation task, making this refined embedding more suitable for the base task. However, this improvement is minimal compared to the overall gain achieved by the guided embedding refinement method, emphasizing the critical role of guided embeddings rather than the specific task used for fine-tuning.

\subsection{Interpretability}

In this section, we explain how our method improves the interpretability, as well as the transparency, of the base recommendation system through a case study from the movie dataset. For ease of exposition, we only extract four attributes, including emotional intensity, intellectual challenge, dialogue complexity and story complexity, among all the twelve attributes. We present below in Figure \ref{fig:case-study} the guided embeddings of movie QS4, user SJA and user LWA.

\begin{figure}[!htbp]
    \centering
    {\includegraphics[width=0.9\textwidth]{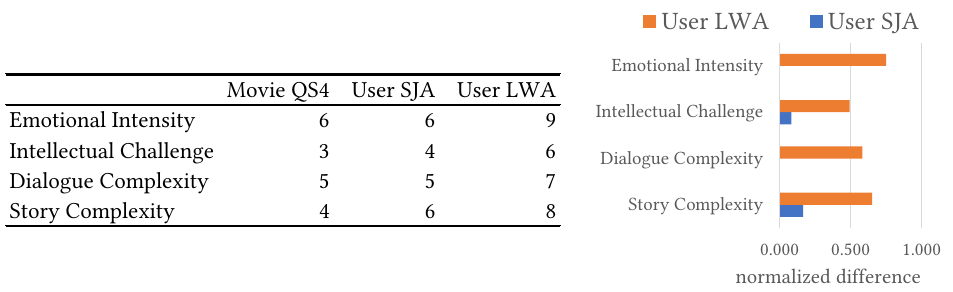}}
    \caption{Aspect-based scores and normalized differnece of movies and users.}
    \label{fig:case-study}
\end{figure}

From the dataset, we know that user SJA expressed a positive opinion (a rating of 5) about this movie, while user LWA gave a negative opinion (a rating of 2). These differing opinions can be interpreted with the help of guided embeddings as follows. From Figure \ref{fig:case-study}, we observe that movie QS4 has moderate emotional intensity and dialogue complexity, with relatively weak intellectual challenge and story complexity. 

The positive feedback from user SJA can be attributed to the fact that user SJA prefers movies with a moderate level across all four attributes. In other words, user SJA's preferences align well with movie QS4. On the other hand, user LWA prefers movies with a moderate level of intellectual challenge and a high level of the remaining attributes. Therefore, user LWA’s preferences do not align as closely with movie QS4 as user SJA's, which explains the negative feedback. Additionally, the guided embeddings for user LWA suggest that we should recommend movies with high levels of emotional intensity, dialogue complexity, and story complexity, in order to better align with user LWA's preferences for these attributes.

\section{Conclusion $\&$ Limitation}
In this work, we propose the guided embedding refinement to improve the performance and interpretability of the base recommendation system. We designed structured queries to prompt LLMs to score items and users based on domain-specific, interpretable attributes, akin to the recommendation logic of an experienced salesperson. The LLMs are fine-tuned to align user preference with item properties on those attributes. We then use the guided embedding and the base embedding with reduced dimension to form the refined embedding. The refined embedding is sent to the base recommendation module for recommendations. Extensive experiments indicate that our method has the following advantages: (1) Generalizability: our approach is adaptable to a variety of recommendation tasks; (2) Performance improvement: the refined embedding boosts recommendation performance by up to $50\%$, even when the base embedding has a higher dimensionality; (3) Enhanced interpretability: our method improves the transparency and understandability of the recommendation system. A limitation of our work is that we have not scaled the model to evaluate performance at a larger scale. The incorporation of more advanced pre-trained LLMs is left as a direction for future work.


\appendix

\section{Implementation Details}

In this section, we provide additional details regarding the implementation of guided embedding refinement.

\subsection{Data Preprocessing}
\label{data;prep}
To ensure data quality and consistency, we follow a standardized preprocessing pipeline, as commonly adopted in sequential recommendation research~\cite{kang2018self, sun2019bert4rec}. Specifically, we conduct the following preprocessing steps:
\begin{enumerate}
    \item  \textbf{Filtering Items}: We remove duplicate entries from the item metadata and filter out items with limited interactions to maintain a representative dataset.
    \item  \textbf{Filtering Users}: We exclude users with an insufficient number of interactions to ensure that enough interaction history can be used to analyze user preferences.
    \item  \textbf{Handling Interaction Records}: We process the user-item interaction logs by removing duplicate entries and sorting interactions chronologically.
\end{enumerate}

\subsection{Aspect Selection for Guided Embedding}

The aspects used in guided embedding generation were manually selected to align with domain-specific characteristics that are meaningful for recommendation. The specific aspects and their descriptions are listed in \Cref{tab:emb-desc-cloth,tab:emb-desc-game,tab:emb-desc-movies}.

\subsection{Finetuning Process}

For fine-tuning the Llama 3.2 1B model on the recommendation task, we employ a reinforcement learning-based optimization strategy using policy gradient methods. The model is fine-tuned separately for each dataset category (Movies, Clothing, and Games) for a total of 12 epochs with a batch size of 4. The learning rate is set to 1e-5 with a $2\%$ warmup ratio, followed by a linear decay schedule. Training is conducted using the Adam optimizer, and mixed-precision training (fp16) is applied to improve computational efficiency and numerical stability. All experiments are executed on an NVIDIA Tesla A100 GPU.

In the reinforcement learning process, rewards are computed at the token level within $\mathcal{L}_{\mathrm{rec}}$.
For tokens that do not correspond to actual embedding values, the system verifies their format correctness.
For tokens that correspond to actual embedding values, rewards are computed based on the model’s performance in the recommendation task.
However, if the output format is completely incorrect and embeddings cannot be extracted, the model falls back to computing a cross-entropy loss using the output from GPT 4o mini. 
This ensures that the model is incentivized to maintain correct formatting.

\subsection{Base Recommendation System}
\label{app;base}
We provide brief introduction for our base recommendation systems:
\begin{itemize}
    \item \textbf{GRU4Rec}~\cite{hidasi2015session}: GRU4Rec is a pioneering session-based recommendation model that employs Gated Recurrent Units (GRUs) to model user interaction sequences. By leveraging recurrent neural networks (RNNs), GRU4Rec effectively captures short-term dependencies in user behaviors.
    \item \textbf{GRU4Rec$^+$}~\cite{hidasi2018recurrent}: An enhanced version of GRU4Rec, this model incorporates additional architectural modifications, such as improved ranking loss functions and better negative sampling strategies. These enhancements lead to more stable training and superior recommendation performance, especially in long-tail item distributions.
    \item \textbf{SASRec}~\cite{kang2018self}: SASRec is one of the first sequential recommendation models to utilize the self-attention mechanism from transformers. Unlike recurrent models, which process interactions sequentially, SASRec applies self-attention to capture both short- and long-term user preferences efficiently. By modeling dependencies between interactions at various temporal distances, SASRec outperforms traditional RNN-based methods in sequential recommendation tasks.
    \item \textbf{BERT4Rec}~\cite{sun2019bert4rec}: Inspired by the BERT (Bidirectional Encoder Representations from Transformers) model in NLP, BERT4Rec adopts a bidirectional transformer architecture for sequential recommendation. It uses a masked item prediction task, akin to BERT’s masked language modeling, to learn rich, context-aware item representations. This design allows BERT4Rec to capture complex user preferences and improves performance in scenarios with sparse interaction data.
\end{itemize}

\subsection{Normalization of Guided Embeddings}

Since the guided embeddings are derived from a pre-trained LLM and have values ranging from 1 to 10, they require normalization before integration with the base embeddings.
Instead of applying a complex linear transformation, we normalize the guided embeddings by first subtracting their mean and then scaling them so that their standard deviation matches the standard deviation of Xavier initialization.
This ensures that the guided embeddings have a distribution consistent with the base embeddings and do not dominate the learning process.

Mathematically, the transformation is given by:
$$e^g = \frac{e^g - \mu_g}{\sigma_g} \times \sigma_x$$
where $\mu_g$ and $\sigma_g$ represent the mean and standard deviation of the guided embeddings, respectively, and $\sigma_x$ is the standard deviation of Xavier initialization.

\subsection{Refined Embedding Training}
After normalization, the guided embedding is concatenated with the base embedding to form the refined embedding.
For all experiments, the hyperparameter $\mu$ is set to $1$ to balance the weight of the guided embedding in the refined embedding.
The refined embedding is then passed to the recommendation module for training and inference.
The loss function for training the sequential recommendation module is defined as $\mathcal{L}_{\mathrm{ref}}$ in Section~\ref{refined embedding}.

For training the base recommendation models, we adopt the original hyperparameter settings specified in the respective papers.
Specifically, GRU4Rec and GRU4Rec+ are trained for 200 epochs, while SASRec and BERT4Rec are trained for 1000 epochs.
When using only base embeddings, we set the embedding dimension to 60.
However, when incorporating guided embeddings, the base embedding dimension is adjusted to 48 to maintain the overall embedding size at 60, ensuring fair performance comparisons.
All the training process is conducted on the same GPU hardware as used for the guided embedding generation.

\section{Additional Ablation Study for Weight Parameter $\mu$}
\label{app;abl}
\begin{figure}[htbp]
    \centering
    {\includegraphics[width=0.9\textwidth]{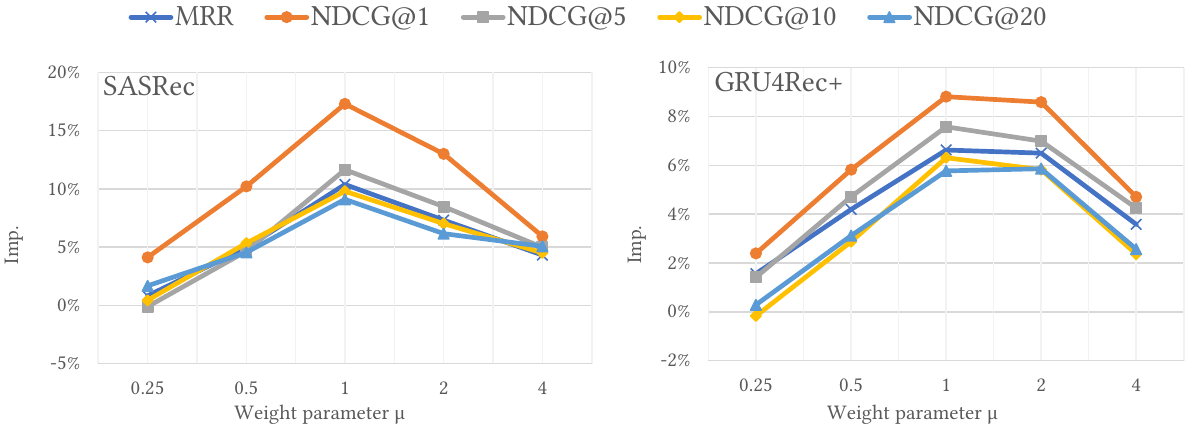}}
    \caption{Impact of weight parameter $\mu$ on performance improvement for SASRec and GRU4Rec$^+$ on the Movies dataset. The x-axis represents different values of $\mu$, while the y-axis indicates the percentage improvement in performance metrics.}
    \label{fig:int_ratio}
\end{figure}

We examine the impact of $\mu$, which controls the weight of guided embeddings in the refined representation, using values of 0.25, 0.5, 1, 2, and 4.
The experimental setup follows that of Section~\ref{sec:int_dim}.
The results on SASRec and GRU4Rec$^+$ for the Movies dataset are visualized in \autoref{fig:int_ratio}.

We observe that the best performance is achieved when $\mu=1$, with consistent improvements in all ranking metrics. Lower values of $\mu$, such as 0.25 and 0.5, lead to weaker gains, suggesting that the weight of the guided embeddings is too light to have an adequate influence on the overall performance. Interestingly, for GRU4Rec$^+$, when $\mu=2$, performance remains stable but does not improve further, indicating a saturation point. However, for $\mu=4$, we observe a slight drop in performance, implying that an excessive weight on guided embeddings might introduce noise or overshadow essential information in the base embeddings.

Notably, the stability of $\mu$-selection can be attributed to the \textbf{normalization} of guided embeddings before constructing the refined representation. By ensuring guided embeddings are normalized, they maintain a balanced magnitude relative to base embeddings. This balancing effect simplifies hyperparameter selection, as it prevents guided embeddings from dominating or being overshadowed by base embeddings.

\section{Prompt Formats}
\label{prompt format}

We list the prompt templates we use in \Cref{tab:guided-item-emb-gen,tab:guided-user-emb-gen}. We also provide several examples of our prompt in \Cref{tab:item-prompt-example,tab:user-prompt-example,tab:review-prompt-example}.

\DefineVerbatimEnvironment{MyVerbatim}{Verbatim}{breaklines=true, commandchars=\\\{\}}

\begin{table*}[t]
    \centering
    \small
    \begin{tabularx}{\linewidth}{lX}
        \toprule
        Aspect & Description \\
        \midrule
        story complexity & How intricate and multi-layered the storyline is (e.g., 1 for very simple, 10 for highly complex). \\
        dialogue complexity & The sophistication of the conversations and monologues (e.g., 1 for simple, 10 for highly sophisticated). \\
        intellectual challenge & How hard to understand the movie is for the audience (e.g., 1 for not challenging, 10 for very hard). \\
        emotional intensity & The degree to which the movie elicits strong emotional responses (e.g., 1 for no emotional impact, 10 for deeply emotional). \\
        visual \& auditory intensity & The impact of visuals, special effects, and sound design (e.g., 1 for minimal impact, 10 for highly intense). \\
        tension level & The suspense and edge-of-your-seat moments (e.g., 1 for no tension, 10 for extremely tense). \\
        pace & The rhythm of the movie, from slow and reflective to fast and action-packed (e.g., 1 for very slow, 10 for very fast). \\
        realism vs. fantasy & Rate from realism to fantasy (e.g., 1 for highly realistic, 10 for pure fantasy). \\
        historical vs. contemporary & Rate from historical to contemporary (e.g., 1 for purely historical, 10 for completely modern). \\
        social value alignment & The degree to which the movie reflects social values or conveys meaningful societal themes (e.g., 1 for no alignment, 10 for strong alignment). \\
        individual viewing vs. group viewing & Rate based on suitability for watching alone vs. with a group (e.g., 1 for perfect for solo viewing, 10 for ideal for group viewing). \\
        movie length (duration) & Evaluate how long is the movie (e.g., 1 for very short, 10 for very long). \\
        \bottomrule
    \end{tabularx}
    \caption{Guided embedding description for Movies dataset.}
    \label{tab:emb-desc-movies}
\end{table*}

\begin{table*}[t]
    \centering
    \small
    \begin{tabularx}{\linewidth}{lX}
        \toprule
        Aspect & Description \\
        \midrule
        color brightness & Bright colors vs. dark colors (e.g., 1 for dark, 10 for bright).\\
        color diversity & Monochromatic vs. multicolored items (e.g., 1 for monochromatic, 10 for multicolored).\\
        complexity & Simple design vs. intricate details (e.g., 1 for simple, 10 for intricate).\\
        shape or structure & Loose vs. fitted designs (e.g., 1 for loose, 10 for fitted).\\
        formality & Casual vs. formal wear (e.g., 1 for casual, 10 for formal).\\
        versatility & Single-purpose vs. multipurpose items (e.g., 1 for single-purpose, 10 for multipurpose).\\
        trendiness & Classic vs. trendy styles (e.g., 1 for classic, 10 for trendy).\\
        social class & Items associated with a social message (e.g., 1 for no social message, 10 for social message).\\
        brand popularity & Well-known vs. niche brands (e.g., 1 for niche, 10 for well-known).\\
        occasion suitability & Everyday use vs. special occasion items (e.g., 1 for everyday, 10 for special occasion).\\
        storage features & Availability of pockets, compartments, or zippers (e.g., 1 for no storage, 10 for multiple storage options).\\
        ease of care & Machine washable vs. dry clean only, need Iron v.s. Iron-free (e.g., 1 for dry clean only, 10 for machine washable).\\
        \bottomrule
    \end{tabularx}
    \caption{Guided embedding description for Clothing dataset.}
    \label{tab:emb-desc-cloth}
\end{table*}

\begin{table*}[t]
    \centering
    \small
    \begin{tabularx}{\linewidth}{lX}
        \toprule
        Aspect & Description \\
        \midrule
        difficulty level & Easy to play vs. challenging to master (e.g., 1 for very easy, 10 for extremely challenging).\\
        genre popularity & Niche genre vs. mainstream genre (e.g., 1 for niche, 10 for mainstream).\\
        game length & Short play sessions vs. long campaigns (e.g., 1 for short, 10 for long).\\
        graphics quality & Retro-style graphics vs. cutting-edge visuals (e.g., 1 for retro, 10 for cutting-edge).\\
        replay value & Low replayability vs. high replayability (e.g., 1 for low, 10 for high).\\
        story depth & Simple storylines vs. intricate narratives (e.g., 1 for simple, 10 for intricate).\\
        multiplayer focus & Solo play vs. multiplayer emphasis (e.g., 1 for solo, 10 for multiplayer).\\
        accessibility & Casual-friendly vs. requiring advanced skills/equipment (e.g., 1 for casual, 10 for advanced).\\
        gaming style and emotional resonance & Relaxed exploration vs. competitive gameplay, Games that elicit strong emotions vs. neutral tones (e.g., 1 for relaxed, 10 for competitive).\\
        customization & Games with deep customization options v.s. limited customization options (e.g., 1 for deep customization, 10 for limited customization).\\
        realism & Preference for realistic simulations vs. fantasy settings (e.g., 1 for realistic, 10 for fantasy).\\
        achievement system & Rich achievement integration vs. basic completion tracking (e.g., 1 for basic, 10 for rich).\\
        \bottomrule
    \end{tabularx}
    \caption{Guided embedding description for Game dataset.}
    \label{tab:emb-desc-game}
\end{table*}

\begin{table*}[t]
    \centering
    \small
    \begin{tabular}{p{\linewidth}}
        \toprule
        \underline{\textbf{\textsc{Prompt Template for Item}}} \\
        \textbf{System Prompt:}
\begin{MyVerbatim}
You are a movie critic. You have been asked to review a movie. 

You should rate the movie on a scale of 1 to 10 (1 is the most negative and 10 is the most positive) on the following dimensions:
{guided embedding aspect}: {guided embedding description}
...
{guided embedding aspect}: {guided embedding description}

Your response should be in the following format, where <rating> is a number between 1 and 10:
{guided embedding aspect}: <rating>
...
{guided embedding aspect}: <rating>

Do NOT include any other information in your response.
\end{MyVerbatim}
\vspace{1ex}
        \textbf{User Prompt:} 
\begin{MyVerbatim}
Here is some information about a movie:
{item prompt}
\end{MyVerbatim}
\\\bottomrule
    \end{tabular}
    \caption{Prompt template for guided item embedding generation.}
    \label{tab:guided-item-emb-gen}
\end{table*}

\begin{table*}[t]
    \centering
    \small
    \begin{tabular}{p{\linewidth}}
        \toprule
        \underline{\textbf{\textsc{Prompt Template for User}}} \\
        \textbf{System Prompt:}
\begin{MyVerbatim}
You are a movie recommender. You will be given some previous movie reviews of a user. You have been asked to recommend a movie to the user.

You should rate the movie which you think the user will like on a scale of 1 to 10 (1 is the least likely and 10 is the most likely) on the following dimensions:
{guided embedding aspect}: {guided embedding description}
...
{guided embedding aspect}: {guided embedding description}

Your response should be in the following format, where <rating> is a number between 1 and 10:
{guided embedding aspect}: <rating>
...
{guided embedding aspect}: <rating>

Do NOT include any other information in your response.
\end{MyVerbatim}
\\
        \textbf{User Prompt:}
\begin{MyVerbatim}
Here are some previous movie reviews of the user:
{user prompt}
\end{MyVerbatim}
        \\\bottomrule
    \end{tabular}
    \caption{Prompt template for guided user embedding generation.}
    \label{tab:guided-user-emb-gen}
\end{table*}

\begin{table*}[t]
    \centering
    \small
    \begin{tabular}{p{\linewidth}}
        \toprule
        \underline{\textbf{\textsc{Example of Item Prompt}}}
\begin{MyVerbatim}
title:
  Escape Plan [DVD + Digital]
description:
  Action superstars Sylvester Stallone and Arnold Schwarzenegger team up in the action-thriller THE TOMB. Ray Breslin (Stallone), the world's foremost authority on structural security, agrees to take on one last job: breaking out of an ultra-secret, high-tech facility called 'The Tomb.' But when he is wrongly imprisoned, he must recruit fellow inmate Emil Rottmayer (Schwarzenegger) to help devise a daring, nearly impossible plan to escape from the most protected and fortified prison ever built.
rating:
  4.5/5.0 (10815 users)
categories:
  Movies & TV
  Genre for Featured Categories
  Action & Adventure
details:
  Genre: Action
  Format: Multiple Formats, Closed-captioned, Color, NTSC, Widescreen
  Contributor: Arnold Schwarzenegger, Jim Caviezil, Sylvester Stallone, Mickael Hafstrom
  Language: English
\end{MyVerbatim}
        \\\bottomrule
    \end{tabular}
    \caption{Example of Item Prompt.}
    \label{tab:item-prompt-example}
\end{table*}

\begin{table*}[t]
    \centering
    \small
    \begin{tabular}{p{\linewidth}}
        \toprule
        \underline{\textbf{\textsc{Example of User Prompt}}}
\begin{MyVerbatim}
review 1:
  {review prompt}
review 2:
  {review prompt}
review 3:
  {review prompt}
...
\end{MyVerbatim}
        \\\bottomrule
    \end{tabular}
    \caption{Example of User Prompt.}
    \label{tab:user-prompt-example}
\end{table*}

\begin{table*}[t]
    \centering
    \small
    \begin{tabular}{p{\linewidth}}
        \toprule
        \underline{\textbf{\textsc{Example of Review Prompt}}}
\begin{MyVerbatim}
rating:
  4.0
reviewTime:
  2014-05-08
summary:
  things i bought from amazon.
reviewText:
  foam cleaner for carpet was not good for me.shampooer was good for price.movie was good. and good too.thank you bye
product:
  {item prompt}
\end{MyVerbatim}
        \\\bottomrule
    \end{tabular}
    \caption{Example of Review Prompt.}
    \label{tab:review-prompt-example}
\end{table*}

\end{document}